\documentclass[twocolumn,letter]{jpsj2} 
\title{Antiferromagnetic fluctuations in Fe(Se$_{1-x}$Te$_{x}$)$_{0.92}$($x$ = 0.75, 1) observed by inelastic neutron scattering}

\author{Satoshi \textsc{Iikubo}$^{1}$\thanks{Present address: Kyushu Institute of Technology, Kitakyushu, Fukuoka 808-0196, Japan},
 Masaki \textsc{Fujita}$^{2}$, 
Seiji \textsc{Niitaka}$^{3}$, 
and Hidenori \textsc{Takagi}$^{3}$ 
}

\inst{$^{1}$WPI Advanced Institute for Materials Research, Tohoku University, Sendai, Miyagi 980-8577, Japan \\
$^{2}$ Institute for Materials Research, Tohoku University, Sendai, Miyagi 980-8577, Japan
\\
$^{3}$RIKEN (The Institute of Physical and Chemical Research), Wako, Saitama 351-0198, Japan
\\
}

\abst{Motivated by the discovery of superconductivity in F-doped LaFeAsO, we investigated the magnetic fluctuations in a related compound Fe(Se$_{1-x}$Te$_{x}$)$_{0.92}$($x$ = 0.75, 1) using neutron scattering techniques. Non-superconducting FeTe$_{0.92}$ shows antiferromagnetic (AF) ordering with the ordering vector $|Q|$ = 0.97 {\rm \AA}$^{-1}$ (Q = (0.5,0,0.5)) at $T_{\rm N} \sim$ 70 K and a structural transition between tetragonal and monoclinic phases. In the AF monoclinic phase, the low-energy spectral weight is suppressed, indicating a possible gap formation. On the other hand, in the paramagnetic tetragonal phase, we observed a pronounced magnetic fluctuation around at $|Q| \approx$ 0.92 {\rm \AA}$^{-1}$, which is slightly smaller than the commensurate value. In Fe(Se$_{0.25}$Te$_{0.75}$)$_{0.92}$, which does not show magnetic ordering and shows superconductivity at $T_c \approx$ 8 K, we observed that a magnetic fluctuation is located at $|Q| \approx$ 0.9 {\rm \AA}$^{-1}$ at low energies and shifts to a higher value of $|Q| \approx$ 1.2 {\rm \AA}$^{-1}$ at higher energies. The latter value is close to a reciprocal lattice vector $Q$ = (0.5,0.5,0.5) or (0.5,0.5,0), where the AF fluctuations are observed in other FeAs-based materials. The existence of this common characteristic in different Fe-based superconductors suggests that the AF fluctuations may play a important role in superconductivity.} 

\kword{Fe(Se$_{1-x}$Te$_{x}$)$_{y}$, magnetic fluctuation, neutron scattering, iron-based superconductor}

\begin{document}
\maketitle


The discovery of superconductivity in the F-doped oxypnictide LaFeAsO with a superconducting transition temperature of $T_{\rm c}$ = 26 K has led to intensive studies on related superconductors including iron group\cite{Kamihara}.
Soon after the discovery, the oxygen-free iron pnictides Ba$_{1-x}$K$_x$Fe$_2$As$_2$ were found to be superconductors with $T_{\rm c}$=38 K \cite{Rotter1,Rotter2}. 
These pnictides possess a FeAs layer in their structure. 
Furthermore, their non-doped counterparts commonly exhibit an antiferromagnetic (AF) ordering with an adjacent structural phase transition from tetragonal to orthorhombic structure.
Since the discovery of superconductivity in these FeAs systems, another new family of Fe-based superconductors, iron chalcogenides Fe(Se$_{1-x}$Te$_{x}$)$_{y}$, has been reported\cite{Hsu, Margadonna,Fang}.
This binary system has drawn considerable attention because of the apparent simplicity of its crystal structure.
These compounds have an $\alpha$-PbO type structure, which consists of a stacking of FeSe(Te) sheets along $c$-axis.
Each sheet consists of a square planar layer of Fe, which is tetrahedrally coordinated to Se(Te), similar to the structure of the FeAs sheets in the above-mentioned pnictides. 
AF ordering is observed at $T_{\rm N}$ = 65 K in end compound FeTe$_{0.82}$\cite{Fang}.
The magnetic ordering disappears and the superconductivity indeed sets in as Se is substituted for Te. 

Although the AF ordering in FeTe$_{0.82}$ qualitatively resembles that in other FeAs systems, such as BaFe$_2$As$_2$ \cite{Rotter1}, recent neutron diffraction measurements\cite{Bao,Li} reported that the ordered magnetic structure of Fe(Se$_{1-x}$Te$_{x}$)$_{y}$ is quite different from that of BaFe$_2$As$_2$.
In the FeAs system, a commensurate magnetic order with a propagation vector $Q$ = (0.5,0.5,0.5) was reported, and the magnetic excitation was characterized by the reciprocal lattice point or (0.5,0.5,0), which corresponds to the 2D nesting vector $Q$ = (0.5,0.5) between the cylinder-like electron and hole Fermi surfaces\cite{Cruz,Qui,Ishikado,Christianson,Ewings,McQueeney,Zhao,Matan}. 
On the other hand, in Fe(Se$_{1-x}$Te$_x$)$_y$, the propagation vector is located at (0.5,0,0.5) for FeTe$_{0.92}$ and the slightly incommensurate (0.38,0,0.5) for FeTe$_{0.82}$\cite{Bao}. 
A precise understanding of the different magnetic properties of the related compounds would provide a useful guideline for understanding the superconductivity. 
So far, however, neutron scattering experiments on Fe(Se$_{1-x}$Te$_x$)$_y$ have been limited to obtaining static information concerning the magnetic properties.
Therefore, we investigated the dynamical magnetic properties of the parent material FeTe$_{0.92}$ and superconductor Fe(Se$_{0.25}$Te$_{0.75}$)$_{0.92}$. 
We studied the evolution of the antiferromagnetic fluctuations as Se was substituted for Te in two samples, and we discuss the possibility of a common magnetic fluctuation with a characteristic 2D wave vector $Q$ = (0.5,0.5) for the superconducting sample. 


We performed elastic inelastic scattering experiments on two polycrystalline samples FeTe$_{0.92}$ and Fe(Se$_{0.25}$Te$_{0.75}$)$_{0.92}$ using the Kinken powder diffractometer HERMES, and the triple-axis spectrometer TOPAN. 
These belong to Tohoku University and are installed at the JRR-3M reactor in the Japan Atomic Energy Agency, Tokai. 
In the HERMES diffractometer, neutrons with a wavelength of 1.8196 {\rm \AA} were obtained by 331 reflection of the Ge monochromator and 12'-blank-sample-18' collimation. 
In the TOPAN spectrometer, the horizontal collimations were open-60'-sample-60'-60'. 
A scattered neutron energy E$_f$ was selected at 13.7 meV was selected using the 002 reflection of pyrolytic graphite.
The powder samples were sealed in an aluminum cell with He gas. The sample cell was mounted at the cold head of a closed-cycle He-gas refrigerator. 




\begin{figure}[t]
\begin{center}
\scalebox{0.75}{\includegraphics{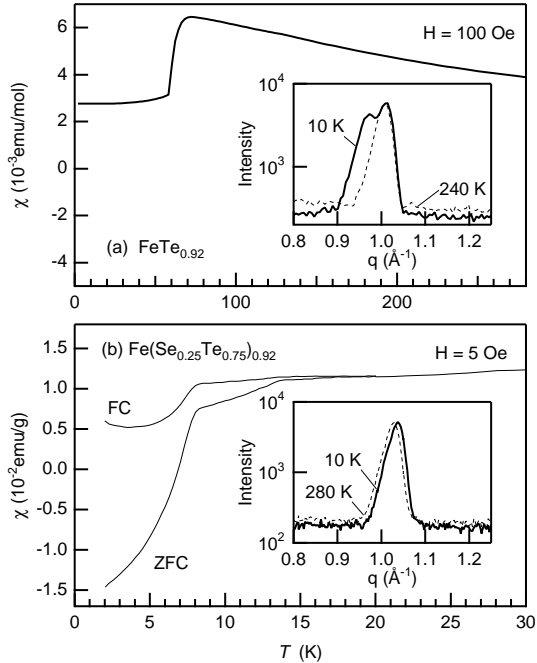}}
\end{center}
\caption{(a) Temperature dependence of magnetic susceptibility for FeTe$_{0.92}$. Inset: Observed neutron powder diffraction data in log scale at 240 K and 10 K. Elastic magnetic intensity at $Q$ = 0.97 {\rm \AA}$^{-1}$ corresponds to the magnetic (0.5,0,0.5) reflection. (b) Zero-field-cooled (ZFC) and field-cooled (FC) susceptibility $\chi$ for Fe(Se$_{0.25}$Te$_{0.75}$)$_{0.92}$. Inset: Observed neutron powder diffraction data in log scale for Fe(Se$_{0.25}$Te$_{0.75})_{0.92}$ at 280 K and 10 K. }
\label{f1}
\end{figure}

\begin{figure}[t]
\begin{center}
\scalebox{0.75}{\includegraphics{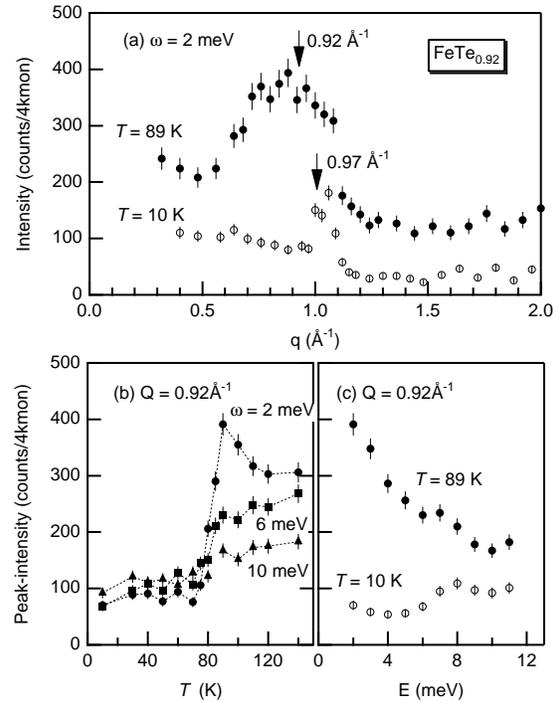}}
\end{center}
\caption{(a) Neutron inelastic scattering intensity of $\omega$ = 2 meV for FeTe$_{0.92}$ at 10 K (open circles) and 89 K (filled circles). (b) Temperature dependence of the peak intensities of the inelastic scattering. Filled circles, squares and triangles represent the data points for $\omega$=2, 6 and 10 meV, respectively. (c) Energy dependence of the peak intensities of inelastic scattering. Open and filled circles represent the data points for {\it T} = 10 K and 89 K, respectively.}
\label{f2}
\end{figure}

Figure 1 (a) shows the magnetic susceptibility of FeTe$_{0.92}$, with the AF ordering appearing as a sharp drop at $\sim$ 70 K.
This anomaly is also associated with a structural transition from a high-temperature tetragonal phase to a low-temperature monoclinic phase.
The inset of Fig. 1 (a) displays the elastic neutron scattering profile of FeTe$_{0.92}$ that was recorded around $Q$ = (1,0,0) at 240 K and 10 K. 
At $T$ = 10 K, we observe superlattice reflections, one of which is located at the low $q$ side of (1,0,0). This observation corresponds to the AF ordering with the ordering vector $Q$ = (0.5,0,0.5).
The ordered moment for Fe is estimated to be $\mu_{Fe}$ = 1.86 $\pm$ 0.02 $\mu_B$ and lies along the b-axis, which is consistent with a previous report\cite{Bao}. 
(Note that a small amount of impurity Fe$_3$O$_4$, which shows the anomaly in $\chi$ at $\sim$ 120 K\cite{Wright}, is observed in our profiles as a second phase.)

Figure 1 (b) shows the magnetic susceptibility $\chi$ for Fe(Se$_{0.25}$Te$_{0.75}$)$_{0.92}$, which shows superconductivity at $T_{\rm c} \approx$ 8 K.
The inset of Fig. 1 (b) shows the elastic neutron scattering profile of Fe(Se$_{0.25}$Te$_{0.75}$)$_{0.92}$ observed at 10 K and 280 K.
No superlattice peaks or peak splitting is observed at low temperatures, which implies that the system involves a tetragonal unit cell without magnetic ordering even at low temperature. 


We first examine the inelastic scattering data for FeTe$_{0.92}$.
Figure 2 (a) shows the {\it q}- dependence of the inelastic scattering of $\omega$ = 2 meV for two different phases. 
For the paramagnetic tetragonal phase at $T$ = 89 K, a strong low-energy magnetic excitation is present in the form of a broad scattering peak centered at $|Q| \approx$ 0.92 {\rm \AA}$^{-1}$.
The peak center is at 0.92 {\rm \AA}$^{-1}$, which is below 0.97 {\rm \AA}$^{-1}$ ($Q$ = (0.5,0,0.5)) or 1.27 {\rm \AA}$^{-1}$ ($Q$ = (0.5,0.5,0.5)) and above 0.82 {\rm \AA}$^{-1}$ ($Q$ = (0.5,0,0)). 
If we estimate the peak center along the (h,0,0.5) line, the value is assigned to an incommensurate $Q$ = (0.47,0,0.5).
At $T$ = 10 K, the spectral weight of the magnetic fluctuations decrease significantly.
We can see a sharp intensity peaked at $|Q| \approx$ 1 {\rm \AA}$^{-1}$, but it arises from a phonon.
The magnetic Bragg peak appears at $|Q| \approx$ 0.97 ${\rm \AA}^{-1}$ in the AF monoclinic phase, as seen in the inset of Fig. 1(a). 
It seems that the structural phase transition from tetragonal to monoclinic phase slightly changes the magnetic propagation vector from $|Q| \approx$ 0.92 {\rm \AA}$^{-1}$ to $|Q| \approx$ 0.97 {\rm \AA}$^{-1}$.
A modification of the magnetic properties caused by the structural transition was also reported for the FeAs system \cite{Ishikado}.
The magnetic excitation in the paramagnetic tetragonal phase is proposed to have a 2D vector $Q$ = (0.5,0.5), while the propagation vector in monoclinic  phase is 3D vector $Q$ = (0.5,0.5,0.5).
A remarkable difference between our Fe(Se$_{1-x}$Te$_x$)$_{0.92}$ system and the FeAs system is the a-b plane component of the propagation vector, which is  $Q_{\rm 2D}$ = (0.5,0) for the former and $Q_{\rm 2D}$ = (0.5,0.5) for the latter. 

The temperature dependence of the peak intensity at $|Q|$ = 0.92 {\rm \AA}$^{-1}$ for three energys around the $T_{\rm N}$ is shown in Fig. 2(b).
Across the structural/magnetic transition, the scattering intensity of $\omega$ = 2 meV decrease rapidly with decreasing {\it T}, while the transition has little effect on the {\it T} dependence of scattering intensity of $\omega$ = 10 meV.
We can see an increase of scattering intensity of $\omega$ = 2 meV as the temperature approaches down to $T_{\rm N}$. 
Although the magnetic transition is first order-like accompanied by a structural transition\cite{Bao,Li},
the pronounced enhancement of the magnetic fluctuations at low energy tells us that this system is close to the AF ordering even in the tetragonal phase, suggesting a large exchange coupling between Fe spins.
In both phases, energy scans were performed for the magnetic signal at $|Q|$ = 0.92 {\rm \AA}$^{-1}$, as shown in Fig. 2 (c). 
A broad diffuse-like scattering takes place above $T_{\rm N}$, while the low-energy spectral weight is suppressed below $T_{\rm N}$. 
At 10 K, an onset of magnetic scattering seems to occur at $\sim$7 meV, indicating possible gap formation.
One of the plausible explanations for the observed energy dependence is the presence of an antiferromagnon with an anisotropic gap, which has been proposed for the pnictide CrSb\cite{Radhakrishna}.
The excitation spectrum resembles that found in CaFe$_2$As$_2$\cite{McQueeney}.

\begin{figure}[tb]
\begin{center}
\scalebox{0.75}{\includegraphics{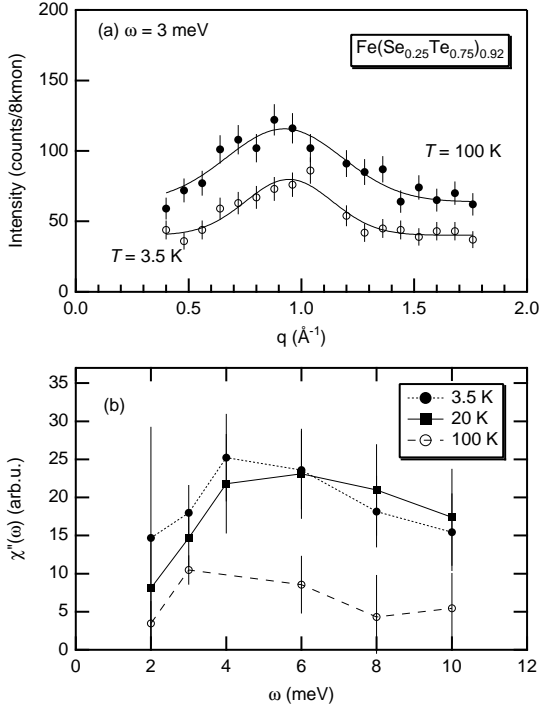}}
\end{center}
\caption{(a) Neutron scattering intensity for Fe(Se$_{0.25}$Te$_{0.75})_{0.92}$ at 3.5 K (open circles) and 100 K (filled circles). (b) Energy dependence of the magnetic excitation spectra $\chi'' (\omega$) at 3.5 K ($T < T_{\rm c}$), 20 K ($T > T_{\rm c}$), and 100 K ($T > T_{\rm c}$).}
\label{f3}
\end{figure}

An interesting question is what happens to the AF magnetic fluctuations when Se is substituted for Te. 
We were able to observe the magnetic fluctuations ($\omega \leq$ 10 meV) over a range from 3.5 to 100 K.
Figure 3 (a) shows the {\it q}- dependence of the inelastic scattering profiles of $\omega$ = 3 meV measured at 3.5 and 100 K. The solid lines represent Gaussian fittings, where the parameters used are the peak center, peak width, scale factor, and background counts that are assumed to be constant with a change in $q$. 
For the sake of clarity, a sharp spurious intensity peak near 1.1 {\rm \AA}$^{-1}$ was subtracted from these profiles. 
A weak but finite broad scattering peak centered at $|Q| \approx$ 0.9 {\rm \AA}$^{-1}$ appears at both 3.5 K and 100 K, indicating that the intensities are almost temperature independent except for the background.
If the scattering intensities arose from a phonons, then the scattering intensities would be expected to be proportionate to the factor $(1-exp(-\hbar \omega/k_BT))^{-1}$, which increases by a factor of $\sim$3.4 from 3.5 K to 100 K. 
Because this increase is not observed in the data, the most likely cause of these peaks is magnetic fluctuations.
Figure 3 (b) shows the energy dependence of the $q$ integrated dynamical magnetic susceptibility $\chi$"($\omega$) below (3.5 K) and above (20 and 100 K) $T_{\rm c}$.
At the lower temperatures, 3.5 and 20 K, the observed $\chi$"($\omega$) show a remarkable enhancement.
Furthermore, the low-energy part of $\chi"(\omega)$ ($\omega < $ 6 meV) increases slightly at 3.5 K.
In the case of FeAs-based materials, the enhancement of the peak intensity of $\chi$"($q$,$\omega$) below $T_{\rm c}$ was interpreted to be a "resonance"\cite{Christianson} as observed in high-$T_{\rm c}$ Cu-oxides; however we could not find the anomalous enhancement in the data.
These data simply seem to indicate that the superconducting gap opening affects the structure of the Fermi surface. 
We think that it is difficult to discuss the relationship between the observed enhancement of $\chi$"($\omega$) and the "resonance" in the absence of other effects on the electronic state near the Fermi surface.


\begin{figure}[t]
\begin{center}
\scalebox{0.75}{\includegraphics{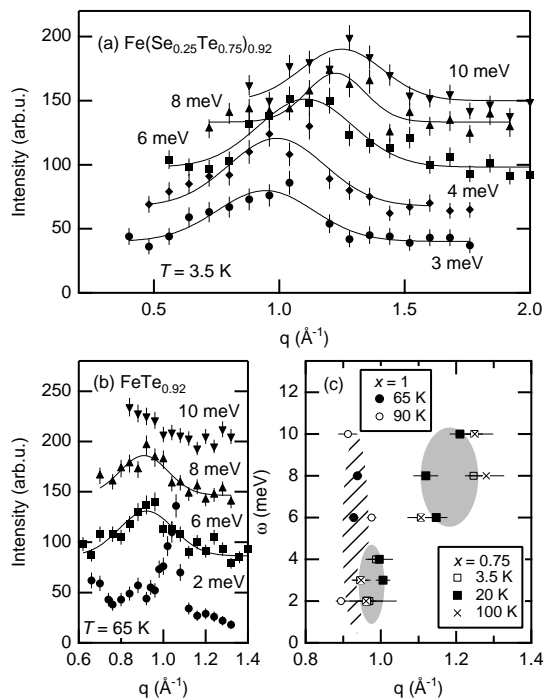}}
\end{center}
\caption{(a) Energy dependence of the inelastic scattering intensity for Fe(Se$_{0.25}$Te$_{0.75})_{0.92}$ at 3.5 K. Solid lines represent the results of fittings the data with a Gaussian line. The zeros of the vertical axis are shifted upwards by 25, 50, 75, and 100 for $\omega$ = 4, 6, 8 and 10 meV, respectively. (b) Energy dependence of the inelastic scattering intensity for FeTe$_{0.92}$ at 65 K. Solid lines represent the results of fittings the data with a Gaussian line. The zeros of the vertical axis are shifted upwards by 50, 100, and 150 for $\omega$= 6 meV, 8meV and 10 meV, respectively. 
(c) Dispersion relation for FeTe$_{0.92}$ at 65 and 89 K and Fe(Se$_{0.25}$Te$_{0.75})_{0.92}$ at 3.5, 20, and 100 K. The lines and the shades are guides to the eye for FeTe$_{0.92}$ and for Fe(Se$_{0.25}$Te$_{0.75})_{0.92}$, respectively.}
\label{f4}
\end{figure}

Next, we will focus on the peak center for Fe(Se$_{0.25}$Te$_{0.75})_{0.92}$.
The energy evolution of the magnetic excitation spectra at 3.5 K is shown in Fig. 4 (a).
For low energies ($\omega \leq$ 4 meV), the magnetic excitations of Fe(Se$_{0.25}$Te$_{0.75})_{0.92}$ peak at a wave vector $|Q| \approx$ 0.9 {\rm \AA}$^{-1}$, which is close to the one for pure FeTe$_{0.92}$.
However, as $\omega$ increase, the peak centers increase substantially up to $|Q| \approx$ 1.2 {\rm \AA}$^{-1}$ at $\omega \approx$10 meV.
Figure 4 (b) shows the energy dependence of magnetic fluctuations for FeTe$_{0.92}$ at 65 K.
Even above 6 meV, the inelastic scattering intensities retain their maximum at $|Q| \approx$ 0.9 {\rm \AA}$^{-1}$. 
(At 2 meV, the sharp intensity caused by a phonon at $|Q| \approx$ 1.0 {\rm \AA}$^{-1}$makes it is difficult to recognize the magnetic signal.) 
Figure 4 (c) shows the peak centers for both samples at several temperature collected at 2 meV $\leq \omega \leq$ 10 meV. 
In case of FeTe$_{0.92}$, the peak centers are almost constant as the function of $\omega$, which is consistent with what has been reported for other FeAs based materials\cite{Cruz,Qui,Ishikado,Christianson,Ewings,McQueeney,Zhao,Matan}.
For Fe(Se$_{0.25}$Te$_{0.75}$)$_{0.92}$, the scattering intensities are located at $|Q| \approx$ 0.9 {\rm \AA}$^{-1}$ below $\omega$= 4 meV. 
Above $\omega \approx$ 6meV, the peak intensity lies at $|Q| \approx$1.2 {\rm \AA}$^{-1}$. This trend is nearly temperature independent.
If we use a spin-wave model, we would expect the high-energy data to exhibit another intensity in the low $q$ side. This simple picture is different from the experimental results. 
In any case, the measurements reported here show conclusively that a significant evolution occurs in the magnetic excitation spectra when Se is substituted for Te.




At this point, we would like to note that the reciprocal lattice point $|Q| \approx$ 1.2 {\rm \AA}$^{-1}$ is close to the wave vectors $Q$ = (0.5,0.5,0) ($|Q|$ = 1.17 {\rm \AA}$^{-1}$) and $Q$ = (0.5,0.5,0.5) ($|Q|$ = 1.28 {\rm \AA}$^{-1}$), which match with the 2D nesting vector between the cylinder-like electron and hole Fermi surfaces like FeAs system. 
One plausible explanation for the peak shift caused by Se substitution is the appearance of the itinerant character of Fe magnetic moment from the parent material FeTe$_{0.92}$. 
Our experimental results on FeTe$_{0.92}$ show that the strong AF correlation with $|Q|$ = 0.92 {\rm \AA}$^{-1}$ leads to a system of AF ordering in three dimensions, accompanied by a structural transition.
Unlike the FeAs-based materials, the most prominent spin fluctuations in FeTe$_{0.92}$ are not $Q$ = (0.5,0.5,0.5) but are instead $Q$ = (0.5,0,0.5) or else have an incommensurate wave vector. 
There are several theoretical explanations that have been offered for the localized spin picture because of the difficulty of explaining $Q$ = (0.5,0.5,0.5) in itinerant electron framework. 
For example, our experimental results may be in agreement with the picture described by F. Ma {\it et al}.\cite{Ma}, which predicted a bi-collinear AF state ($Q$ = (0.5,0,0.5)) with large ordered moments ($\sim$2.5 $\mu_B$) as a ground state of FeTe$_{0.92}$.
The nature of the magnetic properties of the parent material FeTe$_{0.92}$ still remains an open question.
On the other hand, the AF correlation between Fe magnetic moments in Fe(Se$_{0.25}$Te$_{0.75}$)$_{0.92}$ decreases and another itinerant character of magnetic correlation, possibly originating in the Fermi surface nesting, may appears.
The LDA band caluculations found that the Fermi surface structure of FeSe and FeTe had cylindrical electron and hole surfaces very similar to that of the FeAs-based superconductors\cite{Subedi}. 
If so, the spin excitation spectrum would resemble that of FeAs-based materials, consistent with what is observed around $\omega \approx$ 10 meV.
Because of the very weak intensity in Fe(Se$_{0.25}$Te$_{0.75}$)$_{0.92}$, it can not be experimentally determined whether the regions $\omega \approx$ 10 meV are continuous or separated.
At the very least, however, our results show that substituting Se for Te may cause the system to possess characteristic magnetic fluctuations with the 2D nesting vector $Q$ = (0.5,0.5).
We note that there is an independent experimental indication of AF fluctuation with a 2D $Q$ = (0.5,0.5) in FeSe$_{0.5}$Te$_{0.5}$\cite{Mook}.

Our results on Fe(Se$_{1-x}$Te$_{x})_{y}$ system may have significant implications on the physics of Fe-based superconductor from the comparison with FeAs system.
The end compounds in both systems show quite different magnetic orderings from each other.
Nevertheless the superconducting materials in both systems have the surprising common character of the proximity to an AF fluctuation with a 2D $Q$ = (0.5,0.5).
This suggests that the mechanism of superconductivity in two systems may share some common features and especially that the AF correlation with 2D $Q$ = (0.5,0.5) may play an important role in the mechanism of superconductivity in Fe-based superconductors.
It is also interesting to see how the magnetic fluctuations change as the energy increases beyond $\omega$=10 meV.
 




The authors thank K. Enoki and K. Ohoyama for their help in the neutron scattering measurements. 
We are grateful to K. Yamada and K. Horigane for useful discussions.
This work was supported in part by a Grant-in-Aid for Young Scientists (B) (21740239) from the Japanese Ministry of Education, Culture, Sports, Science and Technology.




\end{document}